\documentclass[10pt, aps, pra, twocolumn,nofootinbib,floatfix,showpacs,superscriptaddress]{revtex4-2}
\usepackage[utf8]{inputenc}

\usepackage{graphicx,epsfig,color, colordvi, xypic}
\usepackage{pifont,bbding,amssymb,soul,bbm}
\usepackage{amsmath,amsthm, enumerate, mathtools,slashed, enumitem}
\usepackage{hyperref}

\usepackage{graphicx}% Include figure files
\usepackage{dcolumn}% Align table columns on decimal point
\usepackage{bm}% bold math
\usepackage[dvipsnames]{xcolor}
\usepackage{relsize, accents, wasysym}
\usepackage[capitalize]{cleveref}
\usepackage{times}
\usepackage{comment}

\usepackage{tikz-cd}
\tikzcdset{every label/.append style = {font = \small}}

\usepackage{tikz}
\usetikzlibrary{shapes.geometric}
\usetikzlibrary{shapes.arrows}
\usetikzlibrary {arrows.meta}

\hypersetup{
	colorlinks=true,
	linkcolor=CitingColor,
	citecolor=CitingColor,
	urlcolor=CitingColor
}

\DeclareMathAlphabet\mathbfcal{OMS}{cmsy}{b}{n}

\newcommand{\ket}[1]{\ensuremath{|#1\rangle}}
\newcommand{\bra}[1]{\ensuremath{\langle #1|}}
\newcommand{\braket}[2]{\langle #1|#2\rangle}
\newcommand{\proj}[1]{\ket{#1}\!\bra{#1}}
\newcommand{\ketbra}[2]{\ket{#1}\! \bra{#2}}
\newcommand{\vect}[1]{\mathrm{vec}}
\newcommand{\tr}{{\rm tr}}

\newcommand{\I}{\mathcal{I}}

\newcommand{\vecP}{\vec{P}}

\newcommand{\ma}{\mu}
\newcommand{\mb}{\nu}

\definecolor{CitingColor}{rgb}{0,0.3,1}

\newcommand{\Nt}{N_{\text{\scriptsize tot}}}
\renewcommand{\L}{\mathcal{L}}
\newcommand{\MESd}[1]{\text{MES$_{#1}$}}

\definecolor{nblue}{rgb}{0.2,0.2,0.9}

\begin{document}
\title{Bell inequality violations with random mutually unbiased bases} 

\author{Gelo Noel M. Tabia}
\email{gelonoel-tabia@gs.ncku.edu.tw}
\affiliation{Department of Physics and Center for Quantum Frontiers of Research \& Technology (QFort), National Cheng Kung University, Tainan 701, Taiwan}
\affiliation{Physics Division, National Center for Theoretical Sciences, Taipei 10617, Taiwan}
\affiliation{Center for Quantum Technology, National Tsing Hua University, Hsinchu 300, Taiwan}

\author{Varun Satya Raj Bavana}
\affiliation{Department of Physics, Indian Institute of Technology Kharagpur,
Kharagpur, India, 721302}

\author{Shih-Xian Yang}
\affiliation{Hon Hai Quantum Computing Research Center, Taipei 11492, Taiwan}

\author{Yeong-Cherng Liang}
\email{ycliang@mail.ncku.edu.tw}
\affiliation{Department of Physics and Center for Quantum Frontiers of Research \& Technology (QFort), National Cheng Kung University, Tainan 701, Taiwan}
\affiliation{Physics Division, National Center for Theoretical Sciences, Taipei 10617, Taiwan}

\begin{abstract}
We examine the problem of exhibiting Bell nonlocality for a two-qudit entangled pure state using a randomly chosen set of mutually unbiased bases (MUBs). Interestingly, even if we employ only two-setting Bell inequalities, we find a significant chance of obtaining a Bell violation if the two parties are individually allowed to measure a sufficient number of MUBs. In particular, for the case of maximally entangled qutrits and ququarts, our numerical estimates indicate that we can obtain near-guaranteed Bell violation by considering only such Bell inequalities. The case of maximally entangled ququints is similar, albeit the chance of ending up with a successful trial decreases somewhat to approximately  $99.84\%$. Upon a closer inspection, we find that even all these no-violation instances violate some more-setting Bell inequalities. These results suggest that the experimental tests of Bell nonlocality for these higher-dimensional entangled states remain viable even if the two parties do not share a common reference frame.
\end{abstract}
\date{\today}

\maketitle

\section{Introduction}

Quantum theory is a statistical theory of nature: it does not predict a specific result for an experiment. Instead, it gives the probabilities for obtaining various outcomes. In his seminal works, Bell demonstrated that this quantum indeterminism  is inexplicable by any local-hidden-variable model~\cite{Bell64} nor any locally causal model~\cite{Bell04}. In particular, he showed that correlations produced from measurements on certain entangled quantum states in two spacelike separated locations fail to satisfy what we now call Bell inequalities. Correlations that violate Bell inequalities are said to be \emph{Bell nonlocal}~\cite{Brunner:RMP:2014}. 

Traditionally, an experimental demonstration of Bell nonlocality was perceived only as an empirical test of local causality~\cite{Hensen15,Shalm15,Giustina15,Rosenfeld2017}. 
Nowadays, it also serves as a means for the generation of unpredictable random bits~\cite{Colbeck11,Pironio10,Shalm:2021tv,Liu:2021vp}, the sharing of secret keys~\cite{Nadlinger-DIQKD}, and the certification of the correct functioning of quantum devices~\cite{Schwarz:2016vv,Goh2019} (see also Ref.~\cite{Chen:2021tl}). These fascinating prospects are facilitated by the {\em device-independent}~\cite{Scarani12} nature of a Bell test, i.e., the possibility of inferring Bell-nonlocality directly from the observed data, without relying on any assumption about how the devices involved function.

Typically, to realize Bell nonlocal correlations experimentally, one needs to make a deliberate choice of the measurement bases. Hence, the parties performing a Bell experiment will usually have to employ properly calibrated devices and rely on a shared reference frame. These impose nontrivial requirements in any practical implementation, and therefore, any form of relaxation is surely welcome.
To this end, it was first shown in Ref.~\cite{Liang:PRL:2010} that if the parties share a Bell state and if two measurement bases are chosen uniformly by each party from the Bloch sphere, there is a decent chance ($\approx 28\%$) of finding a Bell-violating correlation. Moreover, this can be boosted to about 41\% if each party selects a random pair of mutually unbiased bases (MUBs). 

Later, this result was strengthened in Refs.~\cite{Shadbolt:SR:2012,Wallman2012} to give almost certain (i.e., up to a set of measure zero) 
%Clauser-Horne-Shimony-Holt~\cite{CHSH1969} (CHSH) 
Bell-inequality violation if each party measures a randomly chosen set of three MUBs. More recently, these randomly chosen triads were also considered in Ref.~\cite{FurkanSenel2015} to show that even in a multipartite scenario, there is a decent chance of demonstrating a strong enough Bell violation for the Greenberger-Horne-Zeilinger state, thereby revealing its genuine multipartite nonlocal nature. In fact, using the tools developed in~\cite{Moroder13}, the latest studies in Ref.~\cite{Yang2020} suggest that with three MUBs, one can already certify the genuine multipartite entanglement present in such states with almost certainty. See also Refs.~\cite{Liang:PRL:2010,deRosier2017,deRosier2020} for related work based on arbitrarily chosen observables and Refs.~\cite{Wallman2011,FurkanSenel2015} for using only two MUBs but in a partially aligned scenario. 

What about higher-dimensional quantum states? In this regard,  Fonseca and Parisio~\cite{FonsecaParisio2015:PRA} provided a first set of results for some family of two-qudit entangled states up to dimensions $d\le 4$. Later, Fonseca {\it et al.}~\cite{Fonseca:PRA:2018} extended the analysis to cover several two-qudit entangled pure states up to dimensions $d \le 10$. Specifically, they considered the scenario where each party performs either two or three alternative measurements, and where the measurement bases are sampled either uniformly according to the Haar measure or correspond to those implementable using multiport beam-splitters and phase-shifters. Their two-setting findings show that the probability of violation decreases monotonically with $d$. Although these chances can be boosted significantly to more than $70\%$ by considering three measurement settings, they are clearly nowhere near the desired near-guaranteed violation (see also~\cite{Lipinska:2018ts}). 
%{\blu See also~\cite{FonsecaParisio2015:PRA} for a discussion of probability of violation for $d=2,3,4$.}

In this work, we follow, instead, the approach in Ref.~\cite{Shadbolt:SR:2012} and  investigate the probability of generating Bell-inequality-violating correlations by performing randomly chosen MUBs on maximally entangled two-qudit states. Note that MUBs are known~\cite{Ji:2008,Liang:2009,Armin:wp} to give maximal quantum violation of certain Bell inequalities, including the CHSH Bell inequality.  More generally, MUBs find applications in quantum state estimation~\cite{WoottersFields1989,Ivonovic1981}, quantum key distribution~\cite{BennettBrassard2014:BB84,Bruss1998,CerfGisin2002}, entanglement detection~\cite{Huang2010,SpenglerHuber2012,BaeHiesmayrMcNulty2019}, and quantum error-correcting codes~\cite{Gottesman1996,CalderbankRains1997}. 

We organize the rest of the paper as follows. We begin by discussing the background concepts and tools we use in in~\cref{Sec:Preliminaries}. Our results concerning Bell nonlocal correlations from the measurement of random MUBs on entangled qudits are presented in ~\cref{Sec:Results}. We conclude with a discussion
of some key observations and possible future directions in ~\cref{Sec:Concluding Remarks} while leaving miscellaneous details to the appendixes.

\section{Preliminaries}
\label{Sec:Preliminaries}

\subsection{Correlations from Bell tests}

A Bell test describes a randomized experiment designed to check whether a given set of observations is compatible with any Bell-local model.
In this work, we consider bipartite Bell scenarios where the two parties Alice and Bob have, respectively,  $\ma$ and $\mb$ possible inputs while each input has $k$ possible outputs. We refer to this as the $(\ma,\mb;k)$ Bell scenario. 
Otherwise, if Alice and Bob have the same number of inputs $m=\ma=\mb$, then we call it the $(m;k)$-scenario.

When testing against a specific Bell inequality, a Bell experiment may equivalently be described in terms of a multi-round nonlocal game~\cite{Cleve2004} played by Alice and  Bob against a referee Ted. Before the game starts, Alice and Bob are informed of the possible queries and responses, as well as the conditions for winning a round. The players can decide on any common strategy beforehand but are not allowed to communicate once the game starts.

In each round, Ted distributes the input $x$ to Alice and $y$ to Bob, where $x,y\in\{0,1,\ldots, m-1\}$. The players respond with outputs $a$ and $b$, respectively, where $a,b,\in \{0,1,\ldots, k-1\}$. 
Let $W(x,y|a,b)$ be a function of $x,y,a,b$ that is equal to $1$ if the winning condition is satisfied and $0$ otherwise. 
Ted collects the outputs and declares that Alice and Bob win the round if $W(x,y|a,b) =1$.
In a quantum Bell experiment, some quantum state $\rho$ is prepared and shared between the players who each chooses a set of $m$ measurements with $k$ outcomes. The input they receive determines their measurement choice in each round, and the outcome gives the output. 

The observations in a Bell experiment provide statistical information about the shared resource between Alice and Bob. Formally, a Bell correlation is the joint conditional probability distribution $P(a,b|x,y)$ that characterizes their input-output relations in a Bell experiment. Since no communication is allowed, 
%First, we require that 
the marginal distribution of the output of each party is expected to be independent of the other party's input. This corresponds to the following no-signaling constraints~\cite{Barrett05}:
\begin{align}
\nonumber
 P(a|x) &= P(a|x,y), &\forall\, a,x,y; \\
 P(b|y) &= P(b|x,y), &\forall\, b,x,y,
\end{align}
%These constraints are 
physically motivated by relativistic causality~\cite{PR1994}, which prohibits faster-than-light information transfer.

A correlation is (Bell-)local if it can be expressed as:
\begin{equation}
\label{eq:localCorrelation}
    P(a,b|x,y) = \int d\lambda\, P(\lambda) P(a|x,\lambda) P(b|y,\lambda),
\end{equation}
i.e., there is a set of local variables $\lambda$ with distribution $P(\lambda)$ shared between Alice and Bob that can reproduce the observed correlation without them knowing each other's input.
If a correlation cannot be cast in the form of Eq.~(\ref{eq:localCorrelation}) then we say that it is (Bell-)nonlocal.

It is useful to give a geometric characterization of the set of all possible local correlations for a given Bell scenario, which we shall call the local set $\mathcal{L}$. Firstly, the set $\mathcal{L}$ is compact and convex. Recall that a compact, convex set is given by the convex hull of its extreme points. In this case, the extreme points of $\mathcal{L}$ correspond to local deterministic points, i.e., correlations such that each input pair $(x,y)$ yields exactly one output pair $(a,b)$. Since we only consider Bell scenarios with finite $m$ and $k$,  we have a finite number of local deterministic points, thus $\mathcal{L}$ is also referred to as the local polytope.
If we need to specify the Bell scenario involved, we shall include it as a subscript, such as $\mathcal{L}_{\ma,\mb; k}$.

In quantum theory, we are interested in correlations obtained by Alice and Bob performing local measurements on a shared quantum state. Formally, the shared state is represented by a density operator $\rho$ and the local measurements are each represented by a positive-operator-valued measure (POVM), which is a set of positive semidefinite operators $E_i$ such that $\sum_{i} E_{i} = \mathbb{I}$. This means that we expect quantum statistics to be given by
\begin{equation}
\label{eq:Bornsrule}
    P(a,b|x,y) = \tr\left(\rho E^{(A)}_{a|x}\otimes E^{(B}_{b|y}\right),
\end{equation}
where  $E^{(A)}_{a|x}(E^{(B)}_{b|y})$ denotes Alice's (Bob's) POVM elements. In general a quantum correlation cannot be put in the form of \cref{eq:localCorrelation}, a fact that is manifested by the violation of some Bell inequality.

\subsection{Bell inequalities}

A Bell inequality is a constraint that has to be satisfied by any correlation $\vecP=\{P(a,b|x,y)\}$ compatible with the Bell-locality condition, \cref{eq:localCorrelation}. The latter, in turn, can be derived, e.g., by assuming a local hidden variable model
with hidden parameter $\lambda$, and that the distribution over $\lambda$ is independent of the measurement settings, i.e. $p(\lambda | x,y) = p(\lambda)$ for all inputs $x,y$. 
Consider, in particular, a linear Bell inequality, which can always be expressed in the form 
\begin{equation}
\label{eq:defineBellInequality}
   \I(\vecP):=\sum_{a,b,x,y} \beta_{abxy} P(a,b|x,y) \le I_\mathcal{L},
\end{equation}
where $\beta_{abxy}\in \mathbb{R}$ are Bell coefficients and $I_\mathcal{L}$ is called the local bound. 
For every Bell inequality of Eq.~(\ref{eq:defineBellInequality}), there are many other physically equivalent but mathematically distinct constraints (having different $\beta_{abxy}$) attainable 
by relabeling.
Simple counting shows that for $n$ parties, $m$ inputs, and $k$ outputs,
we can have up to $n!((k!)^{m!})$ versions. However, some relabeling may lead to the same expression so the actual inequivalent versions is often fewer. 

Since the local set $\mathcal{L}$ forms a convex polytope, it admits two different geometric representations: (i) using the convex hull of its vertices or (ii) using the intersection of a finite number of closed half-spaces. These half-spaces are known as
facets of the polytope, each of which lies on some hyperplane specified by a linear inequality. In the case of $\mathcal{L}$, these half-spaces of maximal dimensions are known as facet (defining) Bell inequalities.
In this work, two facet Bell inequalities are especially of relevance. 

The first of these, known as the Clauser-Horne-Shimony-Holt (CHSH)~\cite{CHSH1969} Bell inequality is the only~\cite{Fine.PRL.1982} nontrivial facet Bell inequality for the $(2;2)$ Bell scenario: 
\begin{equation}\label{Eq_CHSH}
    \mathcal{I}^\mathrm{CHSH} :=\sum_{a,b,x,y=0}^{1} (-1)^{a + b + xy}P(a,b|x,y) \le 2.
\end{equation}
Through a procedure known as lifting~\cite{Pironio2005}, one can also find the (lifted) CHSH inequality as a facet inequality for all the more complicated Bell scenarios. For completeness, we refer the readers to~\cite{Jeba:PRR:2019} for a comprehensive study of the maximal quantum violation of lifted Bell inequalities.

In Bell scenarios with two inputs and $k\ge2$ outputs, there is also the family of Collins-Gisin-Linden-Massar-Popescu (CGLMP)~\cite{CGLMP2002} Bell inequalities (see also~\cite{Kaszlikowski:PRA:2002} for the case of $k=3$), which were shown by Masanes~\cite{Masanes2002} to be facets of $\mathcal{L}$.
Explicitly, these inequalities read as:
\begin{subequations}\label{Ineq_CGLMP}
\begin{align}
\nonumber
    &\mathcal{I}^\mathrm{CGLMP}_{k} := \sum_{i=0}^{\lfloor\tfrac{k}{2} - 1 \rfloor} \left[
    P(A_0 = B_0 + i) + P(A_0 = B_1 - i)  \right. \\
\nonumber 
    &\quad + P(A_1 = B_0 - i -1) + P(A_1 = B_1 +i)  \\
    \nonumber
    &\quad - P(A_0 = B_0 -i - 1) - P(A_0 = B_1 + i + 1)  \\
    &\quad \left. - P(A_1 = B_0 + i) - P(A_1 = B_1 - i - 1 ) \right] 
    \le 2,
\end{align}
where 
\begin{equation}
    P(A_x = B_y + i) := \sum_{j=0}^{k-1} P\left(j,j+i\, (\mathrm{mod}\, k)|x,y\right).   
\end{equation}
\end{subequations}
Notice that for $k=2$, inequality~\eqref{Ineq_CGLMP} reduces to inequality~\eqref{Eq_CHSH} after the relabeling $B_1\leftrightarrow B_0$.

\subsection{Mutually unbiased bases (MUBs)}
\label{Sec_MUBs}

Let $\mathcal{E} = \{ \ket{e_i} \}_{i=0}^{d-1}$ and $\mathcal{F} =\{ \ket{f_j} \}_{j=0}^{d-1}$ denote two orthonormal bases in $\mathbb{C}^d$. We say that $\mathcal{E}$ and $\mathcal{F}$ are \emph{mutually unbiased} if for every pair $(\ket{e_i},\ket{f_j})$ obtained by taking one state from each basis, we have that~\cite{DurtEnglertEtal2010}
\begin{equation}
\label{eq:defineMUB}
    \vert\braket{e_i}{f_j}\vert^{2} = \frac{1}{d}.
\end{equation}
In quantum theory, unbiasedness is often associated with the notion of complementarity. More precisely, a pair of observables $P$ and $Q$ is said to be complementary if their values cannot be measured simultaneously. The tradeoff between determining one observable at the expense of its conjugate pair can be quantified using uncertainty relations. Using Shannon entropy, Maasen and Uffink showed that for any two orthonormal bases $\mathcal{E}$ and $\mathcal{F}$~\cite{MaassenUffink1988},
\begin{equation}
\label{eq:entropicUncertainty}
    H(\vec{p}_\mathcal{E}) + H(\vec{p}_\mathcal{F}) \ge -2\log_2 \max_{\ket{e_i},\ket{f_j}}\vert\braket{e_i}{f_j} \vert,
\end{equation}
where $\vec{p}_{\mathcal{E}} (\vec{p}_\mathcal{F})$ denotes the probability distribution obtained when measuring some fixed but arbitrary quantum state $\rho$ in basis $\mathcal{E} (\mathcal{F})$,
i.e., $p_\mathcal{E}(i) = \bra{e_i}\rho\ket{e_i}$. In finite dimensions, the lower bound of inequality~(\ref{eq:entropicUncertainty}) achieves its largest value $\log_2 d$ when $\mathcal{E}$ and $\mathcal{F}$ are mutually unbiased. Moreover, this is optimal~\cite{Ambainis2010}:
if $\rho = \proj{e_i}$ then $H(\vec{p}_\mathcal{E}) = 0$ and $H(\vec{p}_\mathcal{F}) = \log_2 d$.

A simple way to construct a set of MUBs was first described by Weyl~\cite{Weyl1950}. Define the unitary operators
\begin{equation}
    X = \sum_{j=0}^{d-1} \ketbra{j\oplus 1}{j}, \quad 
    Z = \sum_{j=0}^{d-1} \omega^{j}\proj{j},
\end{equation}
where $\oplus$ denotes addition modulo $d$ and $\omega = e^{2\pi i/d}$ is the root of unity. These are sometimes referred to as the shift and phase operators, the qudit generalization of the Pauli $X$ and $Z$ operators for qubits.
If we take the eigenbases of $Z$, $X$, and $ZX$, then they form a mutually unbiased set.
In prime dimensions, this can be extended to a set with $(d+1)$ bases by adding the eigenbases of $ZX^2, ZX^{3}, \ldots ZX^{d-1}$.
The special status of primes comes from the fact that every $\omega^k$ for $k \in \{1,2,\ldots, d-1\}$ is a primitive root of unity if and only if $d$ is prime. 

The largest possible number of bases in a mutually unbiased set is $(d+1)$~\cite{WoottersFields1989}. When this number is saturated, we say that we have a complete set of MUBs. An important open question is how many MUBs can we find in $\mathbb{C}^d$ for a given dimension $d$. In prime power dimensions we know how to construct a complete set~\cite{WoottersFields1989} but the maximal set is unknown in composite dimensions, the smallest case being $d = 6$. In this case, the largest number of MUBs known is three. One simple way to produce three MUBs is to consider the tensor product of qubit and qutrit MUBs~\cite{McNultyWeigert2012}. 
Even though this procedure results in twelve bases for $d=6$, only up to three of them are mutually unbiased.

\subsection{Estimating the probability of Bell violations}

For any given state $\ket{\psi}$, we are interested in determining the chance of both parties finding a randomly chosen set of MUBs that lead to a Bell-inequality-violating correlation $\vecP$. 
 Of course, for the present purpose, the MUBs are chosen independently by each party, i.e., they are unconstrained with respect to the choice of the other party.
To this end, note that the mutually-unbiased nature of a set of MUBs remains intact even after an arbitrary change of basis. Thus, instead of performing a sampling over randomly chosen MUBs for each party, we can equivalently sample randomly chosen unitaries acting on the respective Hilbert spaces. The MUBs in which the measurements are performed can then be evaluated by conjugating a predefined set of MUBs with the randomly sampled unitaries.

Hence, if Alice and Bob employ, respectively, $\ma$ and $\mb$ randomly chosen MUBs, we may define the probability of violation for $\ket{\psi}$ as:
\begin{equation}
	p_{\ma,\mb}^{\ket{\psi}}:=\int {\rm d}\Omega\, f^{\ket{\psi}}_{\ma,\mb}(\Omega)
\end{equation}
where ${\rm d}\Omega$ represents the Haar measure over two independently chosen qudit unitaries and $f^{\ket{\psi}}_{\ma,\mb}(\Omega)$ is an indicator function that returns unity if and only if the unitarily-conjugated MUBs, when acting on $\ket{\psi}$, give rise to a Bell-inequality-violating correlation via \cref{eq:Bornsrule}.

For general $\ma,\mb, k>2$, the evaluation of $f^{\ket{\psi}}_{\ma,\mb}(\Omega)$, however, can be daunting (see, e.g., Ref.~\cite{Lin2022}). Indeed, for any given $\vecP$ corresponding to a sampled pair of unitaries, this amounts to solving the membership problem with respect to $\L_{\ma,\mb; k}$, which has $k^{\ma+\mb}$ vertices and is of dimension $O(\ma\mb k^2)$. For simplicity, and to see how well the results of Ref.~\cite{Shadbolt:SR:2012,Wallman2012} generalize to higher-dimensional maximally entangled states, we shall consider a {\em superset} relaxation of $\L_{\ma,\mb; k}$ formed by lifting~\cite{Pironio2005} only {\em all} the facet Bell inequalities of $\L_{2; k}$ to the $(\ma,\mb; k)$ Bell scenario. Hereafter, we denote this superset of $\L_{\ma,\mb; k}$ by $\L_{2\to \ma,\mb; k}$, or simply $\L_{2\to \ma,\mb}$ when there is no risk in confusing the value of $k$. Then, we shall be concerned, mostly, with the following lower bound on $p_{\ma,\mb}^{\ket{\psi}}$:
\begin{equation}
	p_{2\to\ma,\mb}^{\ket{\psi}}:=\int {\rm d}\Omega\, f^{\ket{\psi}}_{2\to\ma,\mb}(\Omega),
\end{equation}
where $f^{\ket{\psi}}_{2\to\ma,\mb}(\Omega)$ is an indicator function that returns unity if and only if the unitarily-conjugated MUBs, when acting on $\ket{\psi}$, violates a facet Bell inequality from the $(2;k)$ Bell scenario. For comparison, we also discuss lower bound on $p_{2\to\ma,\mb}^{\ket{\psi}}$ that arises from considering only specific facet Bell inequalities $\I$, such as those given in \cref{Ineq_CGLMP}, then:
\begin{equation}
	p_{\ma,\mb}^{\ket{\psi}} \ge p_{2\to\ma,\mb}^{\ket{\psi}} \ge p_{\text{\tiny $\I$}\to\ma,\mb}^{\ket{\psi}}
\end{equation}

 Since we are interested in finding the Bell violation of a two-qudit state $\ket{\psi}$ with $d$-dimensional MUBs, we focus hereafter on Bell scenarios where $k=d$. In this case, let us denote by $\{ \mathcal{M}_i \}$ a complete (or a maximal) set of MUBs known for dimension $d$ (see \cref{Sec_MUBs}). To estimate these probabilities, we first generate $\Nt$ pairs of unitaries sampled uniformly accordingly to the Haar measure. Then, for each of these trials, we let Alice (Bob) perform her (his)  $\ma$ ($\mb$) measurements in the $U_A$ ($U_B$) conjugated MUBs on the shared state $\ket{\psi}$. From the resulting correlation $\vecP$ defined for the $(\ma,\mb; k)$ Bell scenario, we then {``extract"} $\binom{\ma}{2}\binom{\mb}{2}$ correlations for $(2; k)$-scenario and test each of them against either some specific Bell inequality $\I$ (and all its relabelings) or the complete set of facet Bell inequalities for the $(2; k)$ Bell scenario. { In \cref{App_extraction}, we provide an explicit example illustrating how a two-setting correlation is extracted from the a correlation $\vecP$ defined for the $(3,3; 2)$ Bell scenario.} The fraction of trials $f=\frac{N_\mathrm{viol}}{\Nt}$ that leads to a violation of $\I$ or the complete set of Bell inequalities in the $(2; k)$ Bell scenario then gives rise to, respectively, an estimate of  $p_{\text{\tiny $\I$}\to\ma,\mb}^{\ket{\psi}}$  and $p_{2\to\ma,\mb}^{\ket{\psi}}$.
 Finally, we also report the Clopper-Pearson interval (CPI)~\cite{ClopperPearson1934} for each estimate. The CPI provides an estimate in the uncertainty of the probability of violation based on the cumulative probabilities of the binomial distribution. The values shown were obtained using the $\mathsf{binofit}$ function in MATLAB. 
.
 
 Beyond the probability of violation, we are also interested to know how robust these violations are. In the context of $p_{2\to\ma,\mb}^{\ket{\psi}}$, we may quantify this by
 computing, for each of the $\binom{\ma}{2}\binom{\mb}{2}$ { two-setting} correlations $\vecP'$ extracted from $\vecP$, 
 \begin{align}
 \nonumber
     \max\qquad v \ge 0\qquad\qquad & \\ 
     \text{subject to}\quad v\vecP' + (1-v)\vecP_{w} 
     &\in \L_{2;k}, 
     \label{eq.whiteNoiseVisOptimize}
 \end{align}
where $\vecP_{w}\in\mathcal{L}$ is the uniform probability distribution. The optimum of this linear program, denoted by $v^*$, is known as the white-noise visibility of $\vecP
$ with respect to $\mathcal{L}_{2;k}$.  Notice that for any $\vecP'\in\mathcal{L}_{2;k}$, $v=1$ is a feasible solution to the problem and thus the corresponding $v^*\ge 1$. In other words, 
when $v^*< 1$, $\vecP'$ must lie outside $\mathcal{L}_{2;k}$. With some thought, one notices that the smallest such visibility obtained by solving \cref{eq.whiteNoiseVisOptimize} for  all  $\binom{\ma}{2}\binom{\mb}{2}$ subcorrelations $\vecP'$ extractable from $\vecP$ coincides with that obtained by solving \cref{eq.whiteNoiseVisOptimize} with $\vecP$ replacing $\vecP'$ and  $\L_{2\to \mu,\nu;k}$ replacing $\L_{2;k}$.  After all, the effect of testing all these subcorrelations $\vecP'$ with respect to $\L_{2;k}$ is exactly the same as testing $\vecP$ against $\L_{2\to \mu,\nu;k}$.

Similarly, we can also evaluate the white-noise robustness of any such $\vecP'$ with respect to a given Bell inequality $\I$, cf.~\cref{eq:defineBellInequality}, by computing~\cite{Grandjean2012}:
\begin{equation}\label{Eq:vis-CGLMP}
	v^*_\I = \frac{I_\L-\I(\vecP_w)}{\I(\vecP')-\I(\vecP_w)}
\end{equation}
where $\I(\vecP_w)<I_\L$ and $\I(\vecP')$ are, respectively, the Bell value evaluated for white noise $\vecP_w$ and the extracted correlation $\vecP'$. Again, when the inequality $\I$ is violated by $\vecP'$, we have $\I(\vecP')>I_\L>\I(\vecP_w)$, and thus $v^*_\I<1$. In general, for any specific $\I$, we also have $v^*_\I\ge v^*$ since we can see $v^*$ as the value of $v^*_\I$ minimized over all Bell inequalities defined for that Bell scenario.

\section{Random Bell violations from MUBs}
\label{Sec:Results}

In this section, we present our numerical findings for the probability of violation $p_{2\to\ma,\mb}^{\ket{\psi}}$, with $\ket{\psi}$ being quantum states having local Hilbert space dimension $d=2,\cdots,7$. 
We focus predominantly on the maximally entangled two-qudit state $\ket{\Phi_d} = \tfrac{1}{\sqrt{d}}\sum_{j=0}^{d-1} \ket{j}\ket{j}$, 
which we abbreviate as \MESd{d}. For $d=3,4$, we also include, for the purpose of comparison, $p_{\text{\tiny $\I$}\to\ma,\mb}^{\ket{\psi}}$ with $\I$ being the CGLMP inequality defined in \cref{Ineq_CGLMP}.

\subsection{Qubit:  almost certain Bell violation for \MESd{2}}

The case of $d=2$ has already been investigated rather thoroughly in Ref.~\cite{Liang:PRL:2010,Shadbolt:SR:2012,Wallman2012}. Here, we include the result where Alice measures two MUBs while Bob measures three MUBs on \MESd{2}, namely, $p^{\MESd{2}}_{2\to 2,3}=p^{\MESd{2}}_{2\to 3,2}\approx 90.8\%$. For comparison,  note from Ref.~\cite{Liang:PRL:2010} that $p^{\MESd{2}}_{2\to 2,2}=p^{\MESd{2}}_{\text{CHSH}\to 2,2}\approx 41.3\%$ and from Ref.~\cite{Shadbolt:SR:2012,Wallman2012} that $p^{\MESd{2}}_{2\to 3,3} = 100\%$.

\subsection{Qutrit: near-guaranteed Bell violation for \MESd{3}}

\subsubsection{Maximally entangled two-qutrit state}

Our findings for \MESd{3} are summarized in \cref{tab:Qutrit}. Of particular importance  is the observation that when a complete set of MUBs are used by both parties, and if we test against all the Bell inequalities available in the $(2; 3)$ Bell scenario, we can {\em almost always} find a Bell violation based on these randomly chosen MUBs. Indeed, among the $10^6$ trials carried out in this case, we only find nine instances of $\vecP$ that do not lead to the violation of any two-setting three-outcome Bell inequality. Among them, the largest white-noise visibility found is 1.0061, which implies that $p^{\MESd{3}}_{2\to 4,4}\neq 1$. 

%\begin{table*}%[h!]
\begin{table}[h!]
    \centering
    \begin{tabular}{cc|ccc|ccc}
    \hline \hline
         $\ma$ & $\mb$ & $\Nt$ & $p^{\MESd{3}}_{\I_3\to \ma,\mb}$  & $\mathrm{CPI}_{0.05}$ & $\Nt$ & $p^{\MESd{3}}_{2\to \ma,\mb}$  & $\mathrm{CPI}_{0.05}$ \\
    \hline     
        %  $2$     &  $2$  & $10^{6}$ &  $7.67$    & $(7.62, 7.72)$ & $10^{5}$ &  $32.14$   & $(31.85, 32.43)$ \\
        %  $3$     &  $3$  & $10^{6}$ &  $47.01$   & $(46.91, 47.11)$ &  $10^{5}$  & 98.41& $(98.33, 98.49)$ \\
         $2$     &  $2$  & $10^{6}$ &  $7.67$    & $(7.61, 7.72)$ & $10^{6}$ &  $32.056$   & $(31.96, 32.15)$ \\
         $3$     &  $3$  & $10^{6}$ &  $47.01$   & $(46.91, 47.10)$ &  $10^{6}$  & 98.389& $(98.36, 98.41)$ \\
%         $4$     &  $4$  & $10^{6}$ &  $68.24$   & $(68.15, 68.33)$ &  $10^{6}$  &100.00$^*$ & $100-10^{-4}\times(17, 5)$ \\
         $4$     &  $4$  & $10^{6}$ &  $68.24$   & $(68.15, 68.33)$ &  $10^{6}$  &100.00$^*$ & $(99.9983, 99.9996)$ \\
    \hline
        $2$ & $3$ & $10^{6}$  & $21.63$  & 	$(21.55, 21.71)$ & $10^5$ & $71.08$ & $(70.8, 71.36)$ \\
        $2$ & $4$ & $10^{6}$  & $40.55$  & 	$(40.45, 40.65)$ & $10^5$ & $97.7$ & $(97.61,	97.79)$ \\
        $3$	& $4$ & $10^{6}$  & $60.98$  &  $(60.89, 61.08)$ & $10^5$ & $99.999$ & $(99.994, 100)$ \\
         \hline \hline
    \end{tabular}
    \caption{\label{tab:Qutrit}Summary of the various probabilities of violation estimated for \MESd{3}. The first two columns summarize the number of MUBs employed by each party, $\ma$ for Alice and $\mb$ for Bob. The next three columns list the estimate for $p^{\MESd{3}}_{\I_3\to \ma,\mb}$ where a Bell-violation is only witnessed by the CGLMP inequality $\I^\text{CGLMP}_3$ (abbreviated as $\I_3$) and its relabelings, the number of samples $\Nt$ employed used in this estimation, and the respective CPI at confidence level $\alpha = 0.05$. In the last three columns, we give the estimate for $p^{\MESd{3}}_{2\to \ma,\mb}$ where a Bell-violation is witnessed by {\em any} Bell inequality in the $(2;3)$ Bell scenario, the number of samples $\Nt$ employed used in this estimation, and the respective CPI. The entry 100.00$^*$ means that the corresponding $p^{\MESd{3}}_{2\to \ma,\mb}$ is strictly less than 100\%, but with a difference from unity that is too small to be displayed here (see text for details).}  
\end{table}
%\end{table*}

\begin{figure}[h!]
    \centering
    \includegraphics[width=0.98\linewidth]{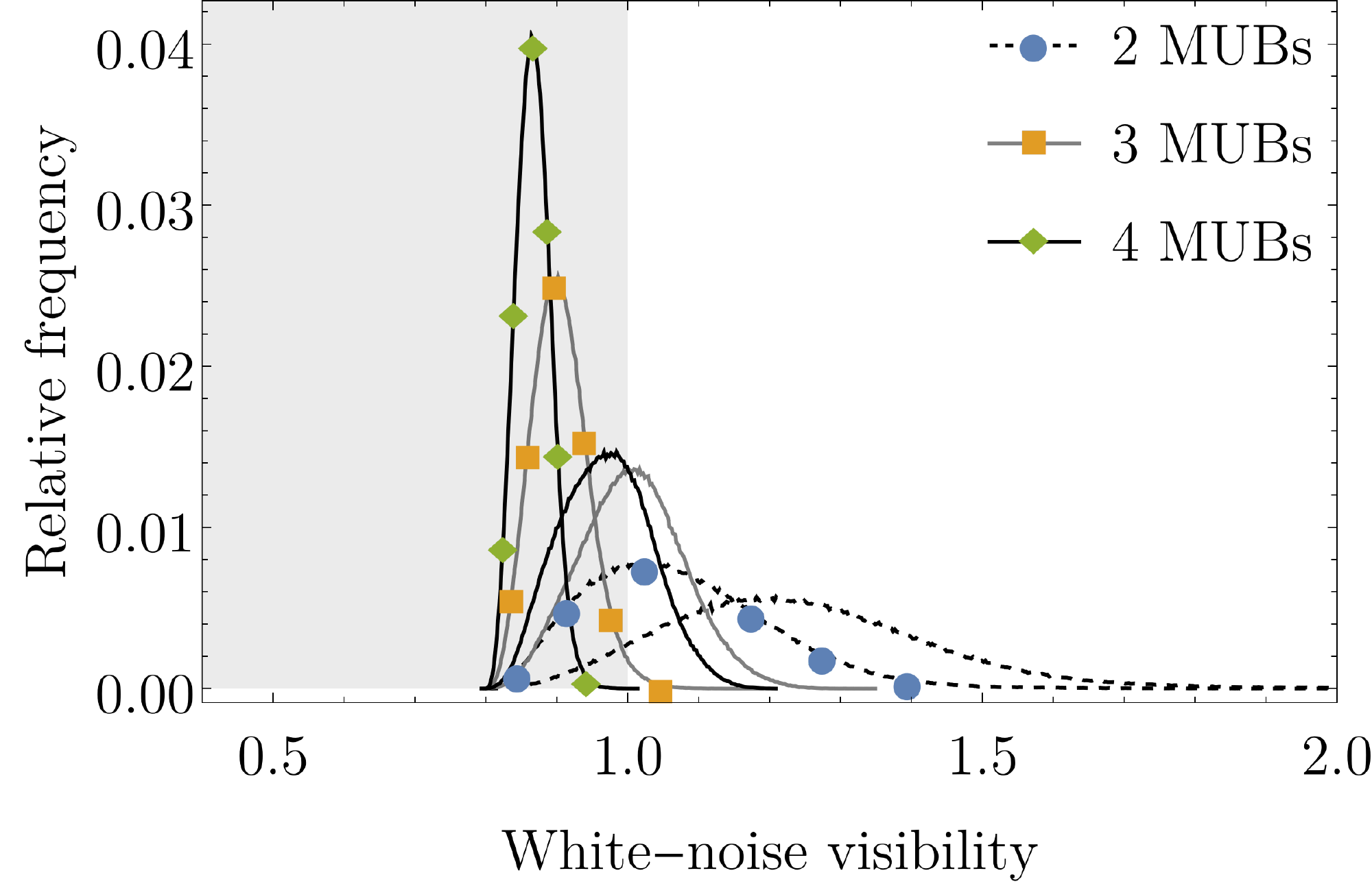}
    \caption{\label{fig:Vis-d3} White-noise visibility histograms for the correlations obtained by performing randomly chosen MUBs on the maximally entangled two-qutrit state. The dashed black line with (blue) circles, solid gray line with (orange) squares, and solid black line with (green) diamonds represent, respectively, the visibility distribution observed during the estimation of $p^{\MESd{3}}_{2\to 2,2}$, $p^{\MESd{3}}_{2\to 3,3}$, and $p^{\MESd{3}}_{2\to 4,4}$, evaluated using \cref{eq.whiteNoiseVisOptimize}; the companion lines without markers are determined using \cref{Eq:vis-CGLMP} based on the CGLMP value for each of the sampled $\vecP$. Here and below (unless otherwise stated), the bin width is set to be $2.5\times10^{-3}$ while the region signifying Bell-nonlocality is shaded in light gray.}
\end{figure}

However, if we test these correlations directly against all the four-setting three-outcome Bell inequalities, we find that they actually lie outside $\L_{4;3}$, giving a white noise visibility of around 0.82 to 0.84. These results suggest that $p_{4,4}^{\MESd{3}}$ could well be unity, i.e., we do end up with  an almost certain Bell violation for \MESd{3} after all. Moreover, as we can see from ~\cref{fig:Vis-d3}, most of the region under the green curve is far away from 1, showing that these randomly generated correlations generally offer a strong white-noise resistance with respect to the Bell polytope of $\L_{2\to 4,4; 3}$.

From~\cref{tab:Qutrit}, it is also clear that the violation of these randomly generated correlation cannot all originate from the violation of a CGLMP inequality. Given that the only other nontrivial class of Bell inequalities~\cite{CollinsGisin2004} for $\L_{2;3}$ is the CHSH~\cite{CHSH1969} Bell inequality lifted~\cite{Pironio2005} to the three-output scenario, it is also natural to wonder how much of these observed violations can be attributed to these lifted inequalities. Our calculation shows that the lifted CHSH on its own can already account for the near-guaranteed violation reported in~\cref{tab:Qutrit}, 
i.e., we have no instances of $\vecP$ that only violate the CGLMP inequality.

\begin{table*}%[h!]
    \centering
    \begin{tabular}{cc|ccc|ccc}
    \hline \hline
         $\ma$ & $\mb$ & $\Nt$ & $p^{\MESd{4}}_{\I_4\to \ma,\mb}$  & $\mathrm{CPI}_{0.05}$ & $\Nt$ & $p^{\MESd{4}}_{2\to \ma,\mb}$  & $\mathrm{CPI}_{0.05}$    \\
    \hline     
$2$ & $2$ & $2\times 10^{4}$ & $0.29$ & $(0.22, 0.38)$ & $10^5$ & $14.74$ & $(14.52,	14.96)$ \\
$3$ & $3$ & $10^{4}$ & $2.25$ & $(1.97, 2.56)$ & $10^5$ & $76.39$  & $(76.13, 76.65)$ \\
$4$ & $4$ & $10^{4}$ & $7.15$ & $(6.65, 7.67)$ & $10^5$ & $99.87$  & $(99.84,	99.89)$ \\
$5$	& $5$ & $10^{4}$ & $12.92$ & $(12.27, 13.59)$ & $5\times10^6$ & $100.00^*$  &  $(99.99989, 100.00^*)$ \\
	\hline					
$2$ & $3$ & $10^{4}$ & $0.79$ & $(0.63, 0.98)$ & $10^4$ & $37.11$ & $(36.16, 38.06)$ \\
$2$ & $4$ & $10^{4}$ & $1.65$ & $(1.41,	1.92)$ & $10^4$ & $62.31$ & $(61.35, 63.26)$ \\
$2$ & $5$ & $10^{4}$ & $2.80$ & $(2.49,	3.14)$ & $10^4$ & $83.74$ & $(83.0, 84.46)$ \\
$3$	& $4$ & $10^{4}$ & $3.82$ & $(3.45, 4.21)$ & $10^4$ & $95.57$ & $(95.15, 95.96)$ \\
$3$ & $5$ & $10^{4}$ & $6.28$ & $(5.81,	6.77)$ & $10^4$ & $99.62$ & $(99.48, 99.73)$ \\
$4$ & $5$ & $10^{4}$ & $10.27$ & $(9.68, 10.88)$ & $10^4$ & $99.99$ & $(99.944, 99.9997)$ \\
         \hline \hline
    \end{tabular}
    \caption{Summary of the various probabilities of violation estimated for \MESd{4}. As with \cref{tab:Qutrit}, the first two columns summarize the number of MUBs employed by each party, the next three columns list the data and results pertinent to the estimate of $p^{\MESd{3}}_{\I_4\to \ma,\mb}$ (with $\I_4$ being a shorthand for the CGLMP inequality $\I^\text{CGLMP}_4$), and the last three columns list the data and results pertinent to the estimate of $p^{\MESd{4}}_{2\to \ma,\mb}$. }
    \label{tab:ququart}
\end{table*}

\subsubsection{Partially entangled two-qutrit state}

One may also wonder the extent to which our observations depend on the state being maximally entangled. To this end, we also consider the following two-parameter family of (partially) entangled 
two-qutrit pure state in the Schmidt basis:
\begin{equation}\label{Eq_NMEState}
    \ket{\alpha,\beta} = \cos\alpha \ket{00} + \sin\alpha\cos\beta \ket{11} + \sin\alpha\sin\beta \ket{22},
\end{equation}
where $\alpha \in (0, \cos^{-1}\tfrac{1}{\sqrt{3}})$ and $\beta \in (0, \tfrac{\pi}{4})$. Note that $\ket{\alpha,\beta}=\ket{\Psi_3}$ is maximally entangled when $\alpha=\cos^{-1}\tfrac{1}{\sqrt{3}}$ and $\beta=\frac{\pi}{4}$. To understand the dependence of the probability of violation on $\alpha$ and $\beta$, we estimate $p^{\ket{\alpha,\beta}}_{2\to 4,4}$ for $31^2=961$ combinations of $(\alpha,\beta)$ by using $\Nt=1000$ trials for each pair.
Here, $\alpha$ and $\beta$ take values, respectively, from $\alpha_i:=\tfrac{i}{30}\mathrm{cos}^{-1}\tfrac{1}{\sqrt{3}}$ and $\beta_j:=\tfrac{j}{30}\tfrac{\pi}{4}$, with $i,j=\{0,1,2,\cdots,30\}$. 

\begin{figure}
    \centering
    \includegraphics[width=0.85\linewidth]{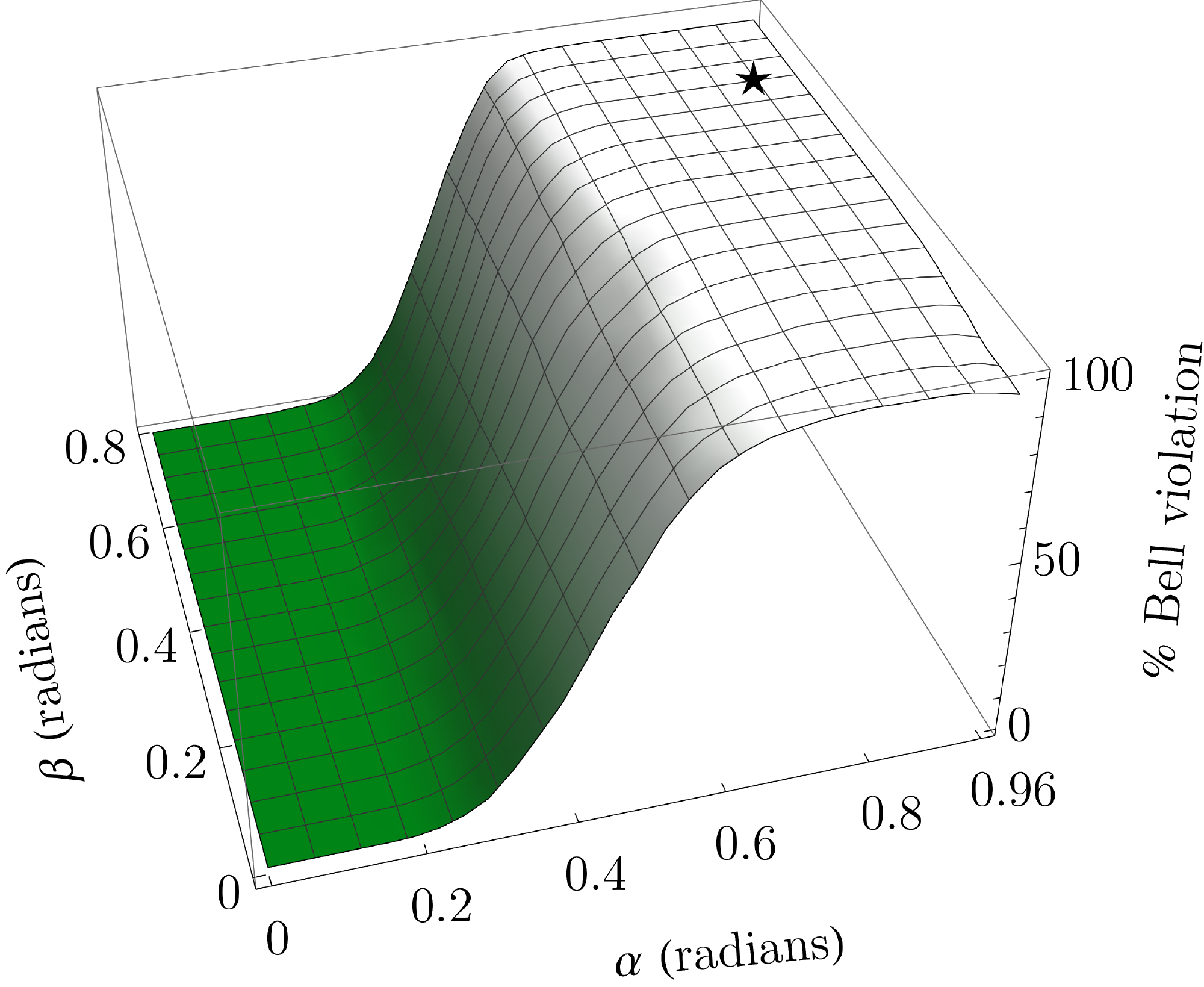}
    \caption{Probability of Bell violation $p^{\ket{\alpha,\beta}}_{2\to 4,4}$ for the two-parameter family of pure two-qutrit entangled states $\ket{\alpha,\beta}$ defined in \cref{Eq_NMEState}. Here, the surface plot is interpolated from 31 equally-spaced values of $\alpha$ from $0$ to $\cos^{-1}\tfrac{1}{\sqrt{3}}$ and 31 equally-spaced values of $\beta$ from $0$ to $\frac{\pi}{4}$ each, making a total 961 combinations of $(\alpha,\beta)$; see text for details. 
     The star indicates the approximate position the nonmaximally entangled state that yields the largest quantum violation of CGLMP.}
    \label{fig:BellViolRandMUBsDim3nonMES}
\end{figure}

From the results shown in \cref{fig:BellViolRandMUBsDim3nonMES}, we see that even though $p^{\ket{\alpha,\beta}}_{2\to 4,4}$ appears to increase with $\beta$, it is predominantly decided by the value of $\alpha$. Specifically, within the fixed interval considered, $p^{\ket{\alpha,\beta}}_{2\to 4,4}$ appears to increase monotonically when $\alpha$ increases. In particular, when $\alpha \approx 0.6687$ (for $i=21$), $p^{\ket{\alpha,\beta}}_{2\to 4,4}$ is already close to unity even though the state is still relatively far from being maximally entangled, since $\cos(0.6687) \approx 0.7846$.

\subsection{Ququart: near-guaranteed Bell violation}

As with the case of qutrits, when a complete set of five MUBs are employed, {\em almost all} the correlations randomly generated from \MESd{4} are provably Bell-nonlocal by using only two-setting four-outcome Bell inequalities, see~\cref{tab:ququart}. In fact, with just four MUBs each for Alice and Bob, we already get $p^{\MESd{4}}_{2\to 4,4}\approx 99.8\%$. However, when all five MUBs are considered, we do find one correlation $\vecP$ among $10^6$ sampled from \MESd{4} that does not violate any two-setting four-outcome Bell inequality. This exceptional correlation $\vecP$ has a white-noise visibility of $1.0005$, thus showing that $p^{\MESd{4}}_{2\to 5,5}\neq 1$.

However, as with the qutrit case, if we test this $\vecP$ against all the five-setting four-outcome Bell inequalities via \cref{eq.whiteNoiseVisOptimize}, we find that it actually lies outside $\L_{5;4}$, giving a white noise visibility of 0.8309. Again, this suggests that $p_{5,5}^{\MESd{4}}$ may indeed be unity.

\begin{figure}[h!]
    \centering
    \includegraphics[width=0.98\linewidth]{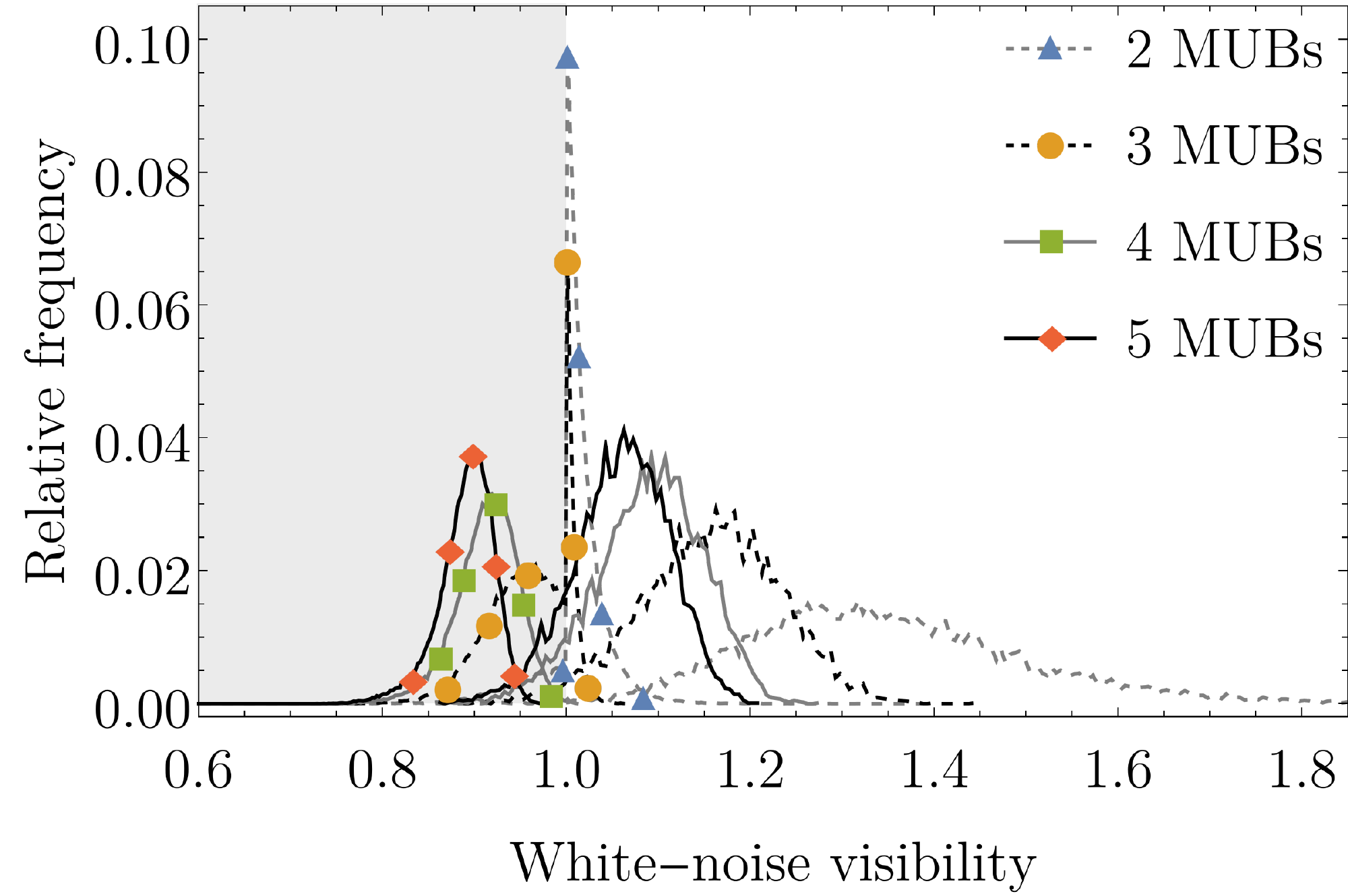}
    \caption{\label{fig:Vis-d4}White-noise visibility histograms for the correlations obtained by performing randomly chosen MUBs on the maximally entangled two-ququart state. The 
    dashed gray line with (blue) triangles, dashed black line with (orange) circles, solid gray line with (green) squares, and solid black line with (dark orange) diamonds represent, respectively, the visibility distribution observed during the estimation of $p^{\MESd{4}}_{2\to 2,2}$, $p^{\MESd{4}}_{2\to 3,3}$, $p^{\MESd{4}}_{2\to 4,4}$, and $p^{\MESd{4}}_{2\to 5,5}$, evaluated using \cref{eq.whiteNoiseVisOptimize}; the companion lines without markers are determined using \cref{Eq:vis-CGLMP} based on the CGLMP value for each of the sampled $\vecP$.}
\end{figure}

Also, as with the two-qutrit case, we see from the solid black line with dark orange diamonds in \cref{fig:Vis-d4} that most of these Bell-nonlocal correlations seem to be fairly robust against the mixture of white noise. 

On the other hand, if we restrict ourself to testing these correlations using only the CGLMP inequalities, then unsurprisingly, the chance of identifying their Bell-nonlocality decreases drastically as compared with the qutrit case, see~\cref{tab:Qutrit} and \cref{tab:ququart}.

\subsection{Ququint: highly-likely Bell violation}

\begin{figure}[h!]
    \centering
        \includegraphics[width=0.95\linewidth]{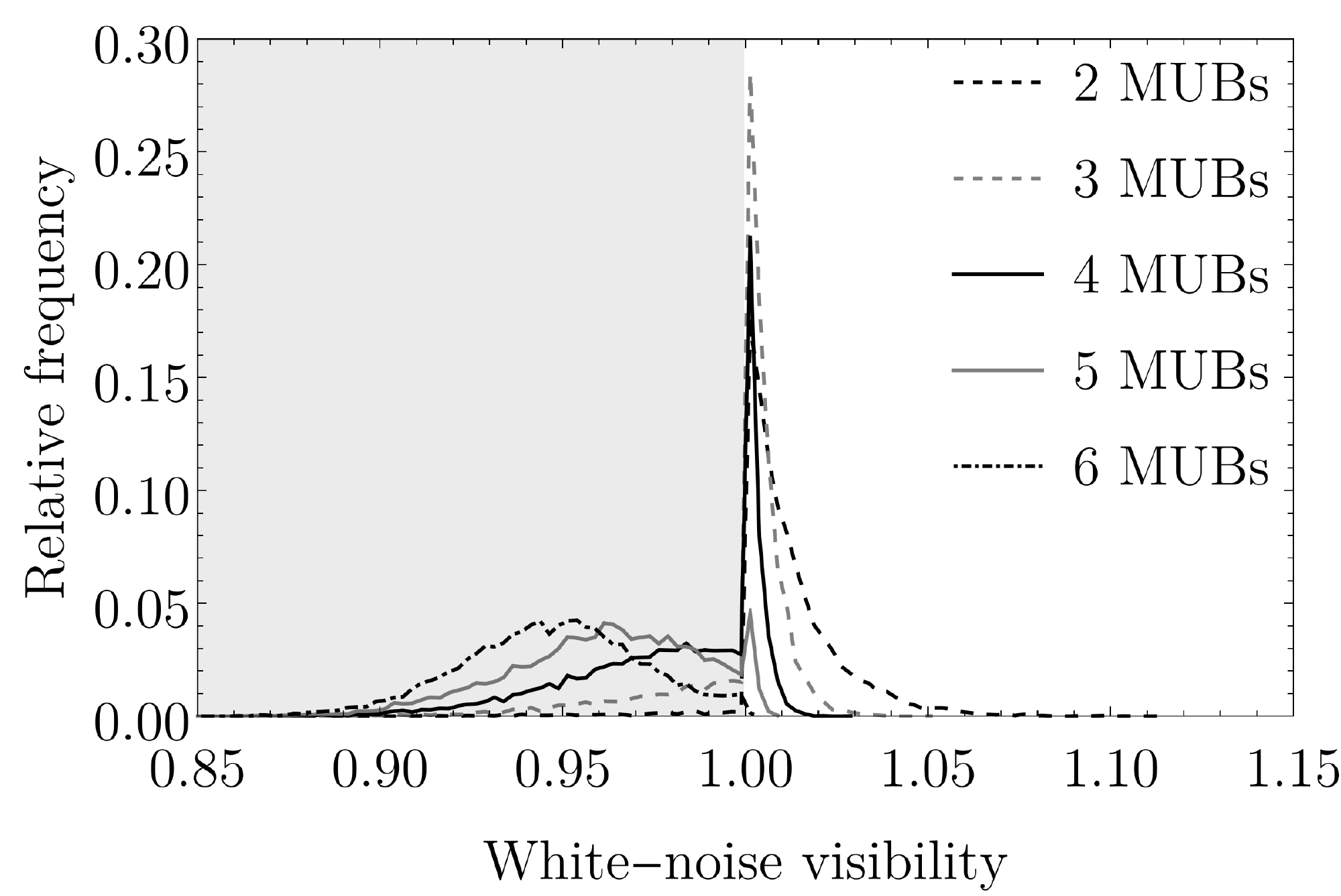}
  \caption{\label{fig:Vis-d5}White-noise visibility histograms for the correlations obtained by performing randomly chosen MUBs on the maximally entangled two-ququint state, cf. \cref{eq.whiteNoiseVisOptimize}. The dashed black line, dashed gray line, solid black line, solid gray line, and dash-dotted black line represent, respectively, the visibility distribution observed during the estimation of $p^{\MESd{5}}_{2\to 2,2}$, $p^{\MESd{5}}_{2\to 3,3}$, $p^{\MESd{5}}_{2\to 4,4}$, $p^{\MESd{5}}_{2\to 5,5}$, and $p^{\MESd{5}}_{2\to 6,6}$.}
\end{figure}

In this case, as can be seen in ~\cref{tab:ququint}, even if a complete set of six MUBs are used to generate the sampled correlations, several instances of the latter are found not to violate any two-setting Bell inequalities. This can also be seen clearly from the corresponding white-noise visibility distribution shown in \cref{fig:Vis-d5}. Specifically, based on $10^4$ trials, our estimate for $p^{\MESd{5}}_{2\to 6,6}$ is $99.84\%$. 

However, as with both the qutrit and ququart case, if more-setting Bell inequalities are allowed, all these correlations are certifiably Bell-nonlocal, see~\cref{tab:ququintNonViol}. This again leaves the possibility of  an almost certain Bell violation for \MESd{5} with MUBs open.

\begin{table}[h!]
    \centering
    \begin{tabular}{ccc|cc}
    \hline \hline
         $\ma$ & $\mb$ & $\Nt$ & $p^{\MESd{5}}_{2\to \ma,\mb}$  & $\mathrm{CPI}_{0.05}$    \\
    \hline     
$2$ & $2$ & $10^4$  & $3.08$  &  $(2.75, 3.44)$ \\
$3$ & $3$ & $10^4$  & $23.28$ & $(22.45,	24.12)$ \\
$4$ & $4$ & $10^4$  & $64.84$ &	$(63.9,	65.77)$ \\
$5$ & $5$ & $10^4$  & $94.00$ & $(93.52, 94.46)$ \\
$6$ & $6$ & $10^4$  & $99.84$ & $(99.74,	99.91)$ \\
\hline
$2$ & $3$ & $10^4$ &  $8.38$ &  $(7.84, 8.94)$ \\
$2$ & $4$ & $10^4$ & $15.33$ &  $(14.63,	16.05)$ \\
$2$ & $5$ & $10^4$ & $25.49$ &  $(24.64, 26.36)$ \\
$2$ & $6$ & $10^4$ & $35.91$ &  $(34.97,	36.86)$ \\
$3$ & $4$ & $10^4$ & $40.81$ &  $(39.85,	41.78)$ \\
$3$ & $5$ & $10^4$ & $57.93$ &  $(56.96,	58.9)$ \\
$3$ & $6$ & $10^4$ &  $73.7$ &  $(72.83, 74.56)$ \\
$4$ & $5$ & $10^4$ & $82.46$ &  $(81.7, 83.2)$ \\
$4$ & $6$ & $10^4$ & $92.39$ &  $(91.85,	92.9)$ \\
$5$ & $6$ & $10^4$ & $98.58$ &  $(98.33,	98.8)$ \\
         \hline \hline
    \end{tabular}
    \caption{Probabilities of violation $p^{\MESd{5}}_{2\to \ma,\mb}$ estimated for \MESd{5}.} 
    \label{tab:ququint}
\end{table}

\begin{table}[h!]
\centering
\begin{tabular}{c|c} 
 \hline
 Smallest visibility $v^*_{(2;5)}$ & Smallest visibility $v^*_{(3;5)}$\\ [0.5ex] 
 \hline\hline
   1.00002 &  0.97848 \\
   1.00008 &  0.97707 \\
   1.00011 &  0.95696 \\
   1.00019 &  0.95161 \\
   1.00031 &  0.94172 \\
   1.00062 &  0.97782 \\
   1.00069 &  0.96014 \\
   1.00071 &  0.96421 \\
   1.00073 &  0.96287 \\
   1.00086 &  0.94129 \\
   1.00088 &  0.95822 \\
   1.00092 &  0.96384 \\
   1.00127 &  0.97256 \\
   1.00165 &  0.96569 \\
   1.00199 &  0.95567 \\
   1.00357 &  0.96540 \\ [0.5ex]
 \hline
\end{tabular}
\caption{White-noise visibilities for the 16 instances of $\vecP$ obtained from \MESd{5} that do not violate any two-setting Bell inequalities. On the left, we list the smallest visibilities $v^*_{(2;5)}$ with respect to only two-setting Bell inequalities (lifted to a six-setting Bell scenario) whereas on the right, we list the corresponding smallest visibilities $v^*_{(3;5)}$ with respect to only three-setting Bell inequalities.}
\label{tab:ququintNonViol}
\end{table}

\subsection{Higher-dimensional cases}

\begin{figure}[h!]
    \centering
        \includegraphics[width=0.85\linewidth]{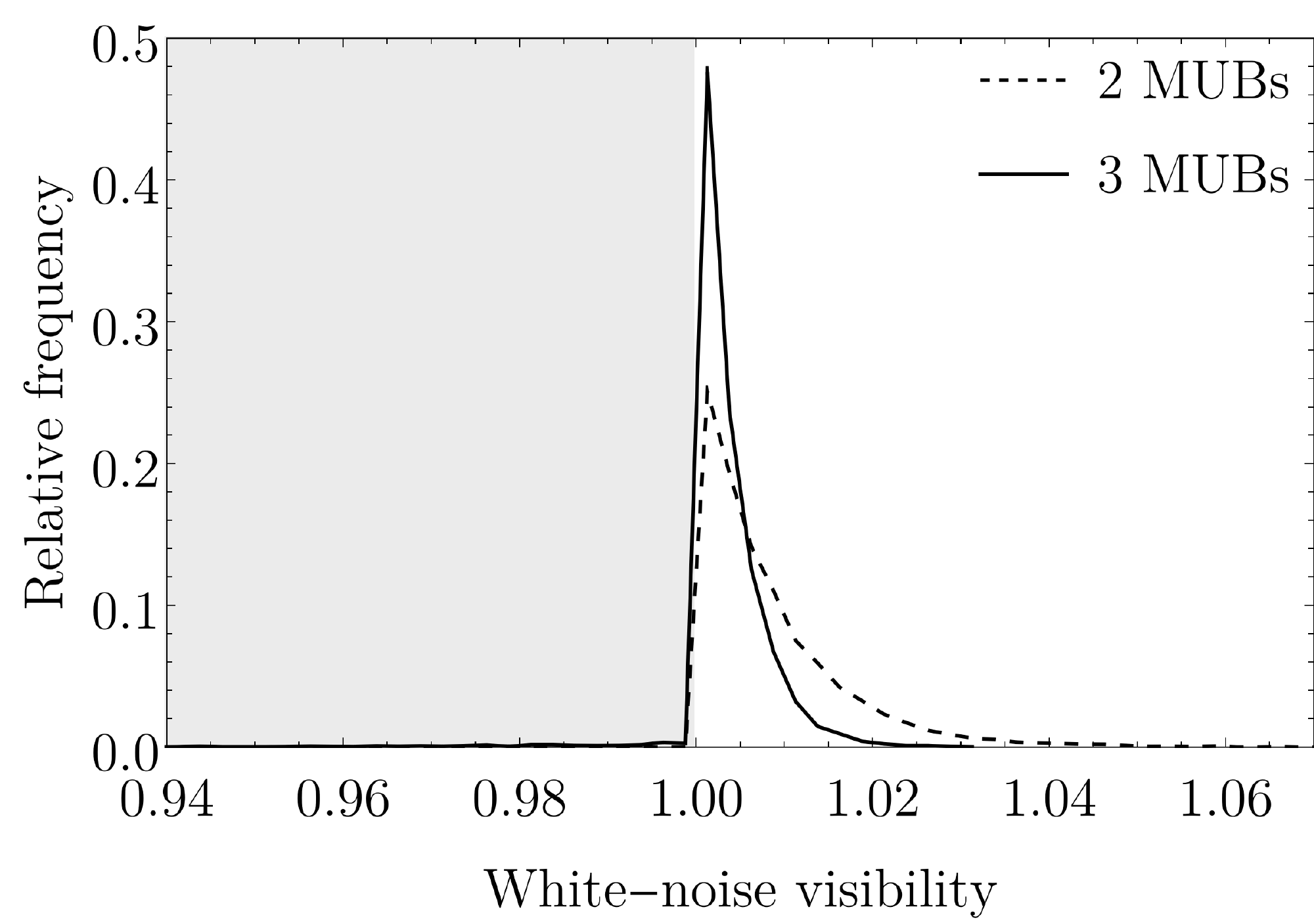}
  \caption{White-noise visibility distribution for the correlations obtained by performing randomly chosen MUBs on the maximally entangled two-qudit state  \MESd{6}, cf. \cref{eq.whiteNoiseVisOptimize}. The dashed and solid black lines correspond, respectively, to the visibility distribution used in the estimation of $p^{\MESd{6}}_{2\to 2,2}$ and $p^{\MESd{6}}_{2\to 3,3}$. The region where the corresponding visibility signifies Bell-nonlocality is shaded in gray.}
    \label{fig:Vis-d6}
\end{figure}

For $d=6$, we  consider only two or three MUBs for each party, see \cref{Sec_MUBs}. In contrast with the $d<6$ cases, our results summarized in \cref{tab:d6} show that the probability of violation $p^{\MESd{6}}_{2\to \ma,\mb}$ is tiny. For completeness, we include a plot showing the visibility distribution in this case in~\cref{fig:Vis-d6}.

\begin{table}[h!]
    \centering
\begin{tabular}{ccc|cc}
   \hline \hline
         $\ma$ & $\mb$ & $\Nt$ & $p^{\MESd{6}}_{2\to \ma,\mb}$  & $\mathrm{CPI}_{0.05}$    \\
    \hline      
$2$ & $2$ &  $10^4$ & $0.28$ &  $(0.19, 0.4)$ \\
$3$ & $3$ &  $10^4$ & $2.77$ &  $(2.46, 3.11)$ \\
\hline											
$2$ & $3$ & $10^4$ & $1.01$ &  $(0.82, 1.23)$ \\
         \hline \hline
    \end{tabular}
    \caption{Probability of violation  $p^{\MESd{6}}_{2\to \ma,\mb}$ estimated for \MESd{6}.} 
    \label{tab:d6}
\end{table}

\begin{figure}
    \centering
 \includegraphics[width=0.95\linewidth]{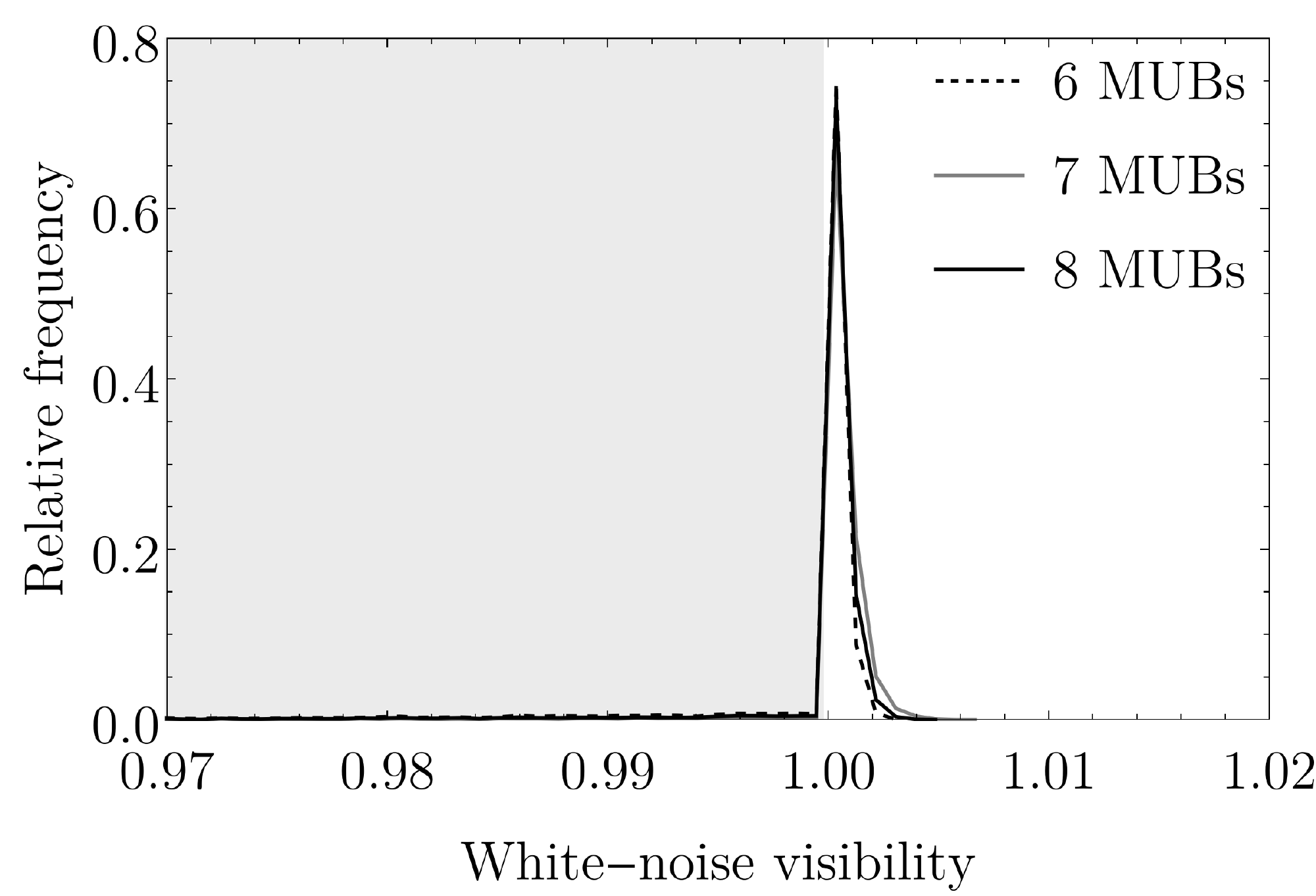}
  \caption{\label{fig:Vis-d7}White-noise visibility histograms for the correlations obtained by performing randomly chosen MUBs on the maximally entangled two-qudit state with $d=7$, cf. \cref{eq.whiteNoiseVisOptimize}. The dashed black line, solid gray line, and solid black line represent, respectively, the visibility distribution observed during the estimation of $p^{\MESd{5}}_{2\to 6,6}$, $p^{\MESd{5}}_{2\to 7,7}$, and $p^{\MESd{5}}_{2\to 8,8}$. The bin width is set to $9\times10^{-4}$ in this case.}
\end{figure}

\begin{figure}[ht]
    \centering
        \includegraphics[width=0.85\linewidth]{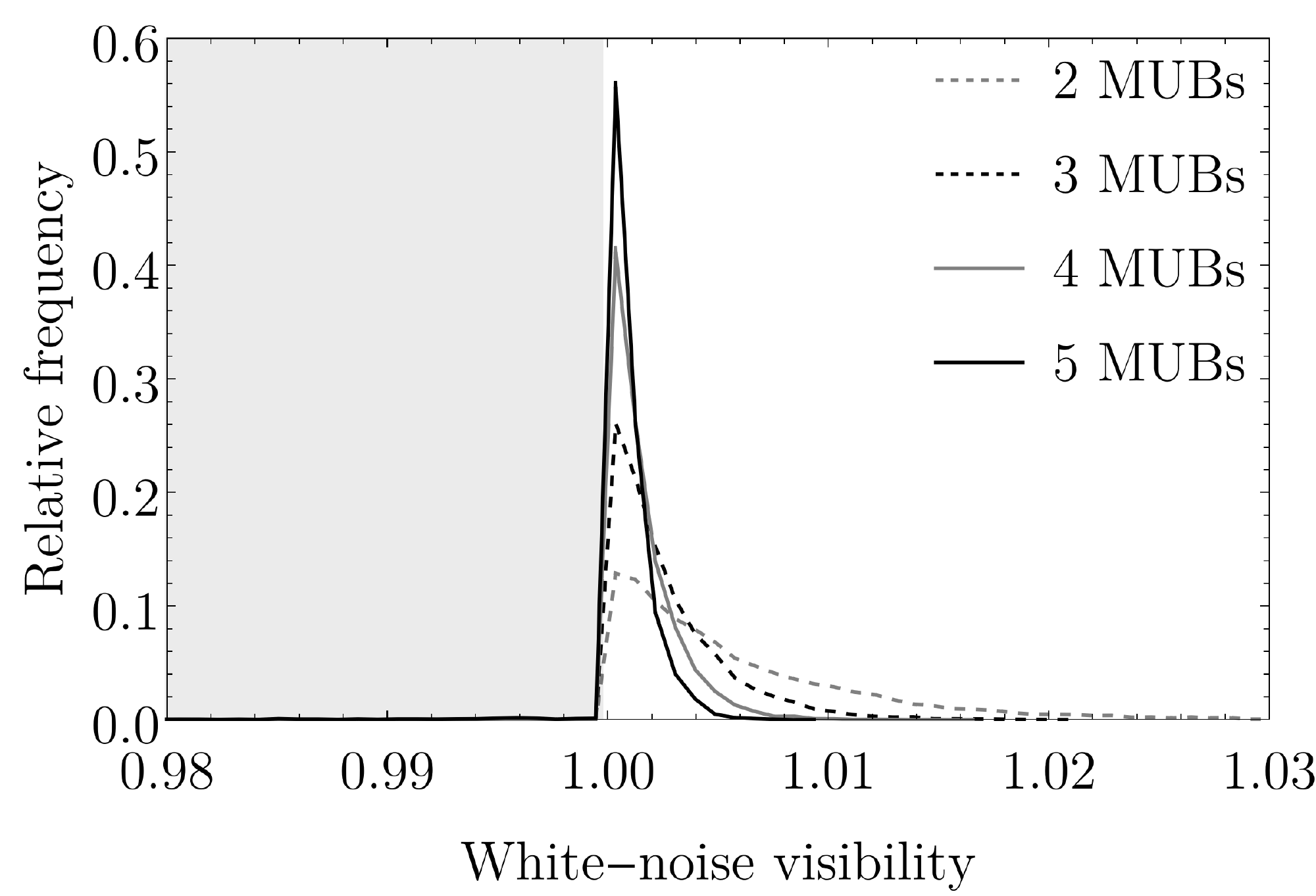}
  \caption{\label{fig:Vis-d7b}White-noise visibility histograms for the correlations obtained by performing randomly chosen MUBs on the maximally entangled two-qudit state \MESd{7}, cf. \cref{eq.whiteNoiseVisOptimize}. The dashed gray line, dashed black line, solid gray line, and solid black line represent, respectively, the visibility distribution observed during the estimation of $p^{\MESd{7}}_{2\to 2,2}$, $p^{\MESd{7}}_{2\to 3,3}$, $p^{\MESd{7}}_{2\to 4,4}$, and $p^{\MESd{7}}_{2\to 5,5}$. The bin width is set to $9\times10^{-4}$ in this case.}
\end{figure}

For $d=7$, although we once again have a complete set of MUBs, the probability of violation $p^{\MESd{7}}_{2\to \ma,\mb}$, as summarized in~\cref{tab:d7}, is substantially smaller than those found for the $d<6$ cases. In fact, even when all eight MUBs are employed, the chance of finding a two-setting Bell-inequality violation from these randomly generated correlations is less than $20\%$. The corresponding visibility distribution is shown in \cref{fig:Vis-d7} when six or more MUBs are considered, and in \cref{fig:Vis-d7b} when fewer MUBs are considered.

\begin{table}[t!]
    \centering
\begin{tabular}{ccc|cc}
   \hline \hline
         $\ma$ & $\mb$ & $\Nt$ & $p^{\MESd{7}}_{2\to \ma,\mb}$  & $\mathrm{CPI}_{0.05}$    \\
    \hline 
$2$ & $2$ & $10^4$ & $0.01$ &  $(0.00025, 0.0557)$ \\
$3$ & $3$ & $10^4$ & $0.15$ &  $(0.084, 0.247)$ \\
$4$ & $4$ & $10^4$ & $0.64$ &  $(0.49, 0.82)$ \\
$5$ & $5$ & $10^4$ & $2.07$ &  $(1.80, 2.37)$ \\
$6$ & $6$ & $10^4$ & $4.87$ & $(4.46, 5.31)$ \\
$7$ & $7$ & $10^4$	& $9.14$ &  $(8.58, 9.72)$ \\
$8$ & $8$ & $10^4$	& $15.5$ &  $(14.80, 16.22)$ \\
         \hline \hline
    \end{tabular}
    \caption{Probabilities of violation $p^{\MESd{7}}_{2\to \ma,\mb}$ estimated for \MESd{7}.} 
    \label{tab:d7}
\end{table}

\section{Concluding Remarks}
\label{Sec:Concluding Remarks}

Motivated by the results of almost certain Bell-inequality violation presented in \cite{Shadbolt:SR:2012} for a Bell state, we investigate the chance of finding a Bell-inequality-violating correlation by performing uniformly and randomly chosen   $d$-outcome MUBs measurements on a two-qudit maximally entangled state $\ket{\Phi_d}$. Specifically, we consider  $d=3,4,5,6$, and $7$, and with both parties employing up to $d+1$ MUBs, if available.

Our results show that even if we consider only two-setting Bell inequalities, as in the case of Refs.~\cite{Shadbolt:SR:2012,Fonseca:PRA:2018}, we can obtain for $d=3$ and $4$ near-guaranteed Bell-inequality violation on $\ket{\Phi_d}$ if we employ the complete set of $(d+1)$ MUBs and optimize over all $\left[\frac{d(d+1)}{2}\right]^2$ pairs of MUBs chosen. Similarly, for $d=5$, when all six MUBs are considered and the same optimization is performed, we end up with $\approx 99.85\%$ chance of finding a correlation that violates a two-setting Bell inequality. Unfortunately, we do not see the same trend when we consider higher-dimensional $\ket{\Phi_d}$. For $d=6$ and $7$, this chance drops, respectively, to below $3\%$ and $16\%$.

\begin{figure}[t!]
    \centering
    \includegraphics[width=0.95\linewidth]{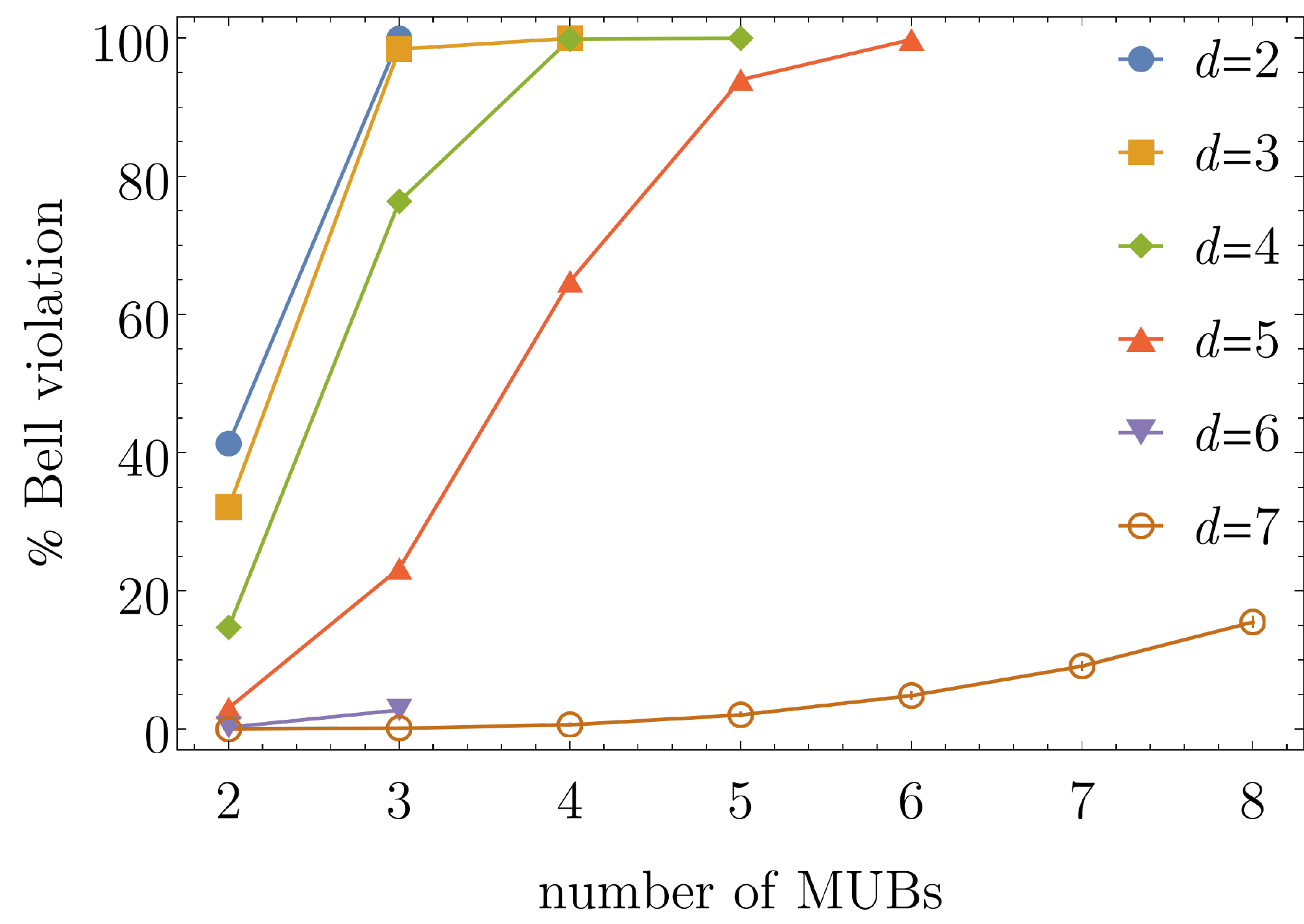}
    \caption{Probability of Bell violation $p^{\MESd{d}}_{2\to \ma,\ma}$ as a function of $\ma$, the number of $d$-outcome MUBs measurements employed by both Alice and Bob on the maximally entangled two-qudit state, for $d=2,3,\ldots, 7$.}
    \label{fig:BellViolRandMUBsSummary}
\end{figure}

More generally, we have observed that for a given $d$, the chance of finding an $\ma$-tuple of Bell-inequality-violating MUBs generally increases with $\ma$, see~\cref{fig:BellViolRandMUBsSummary}. This can be appreciated by noting that as $\ma$ increases, the number two-setting Bell tests that one can perform with the observed correlation increases as $\frac{\ma(\ma-1)}{2}$. This is reminiscent to the observation made in the context of probability of violation for a given state in~\cite{Shadbolt:SR:2012,Fonseca:PRA:2018,Lipinska:2018ts} as well as that found for the sampling of correlations directly from the non-signaling space in~\cite{Duarte_2018,Lin2022}.

\begin{figure}[t!]
    \centering
    \includegraphics[width=0.95\linewidth]{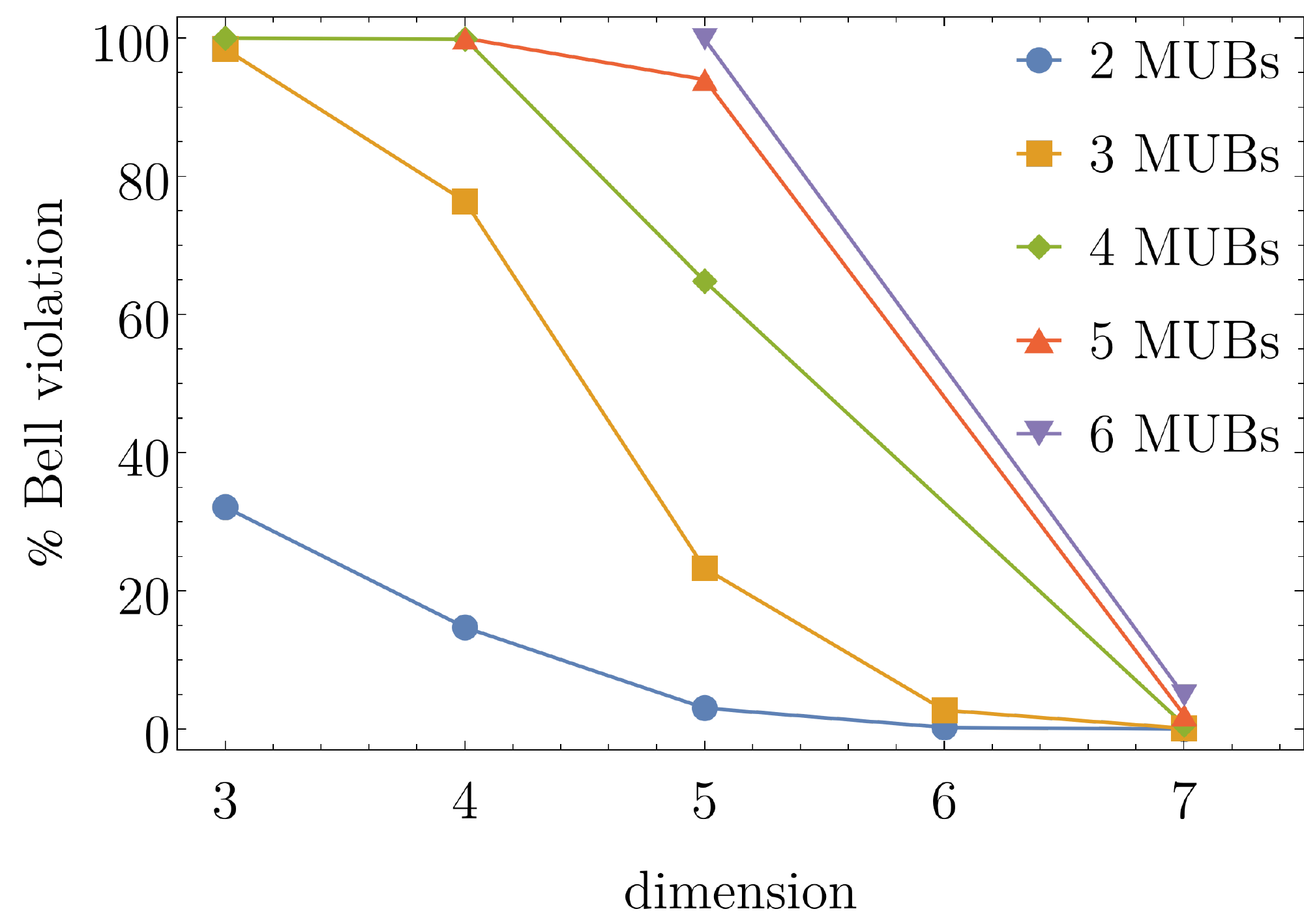}
    \caption{Probability of Bell violation $p^{\MESd{d}}_{2\to \ma,\ma}$ as a function of $d$, the local Hilbert space dimension when the number of MUBs employed are $2,3,\cdots,6$. }
    \label{fig:BellViolRandMUBsTrend}
\end{figure}

On the other hand, if we fix the number of MUBs and increases $d$, a rapidly decreasing trend is observed, see~\cref{fig:BellViolRandMUBsTrend}. Interestingly, the same observation was also found~\cite{Fonseca:PRA:2018} for measurements based on multiport-beam-splitter and phase-shifters. These observations are consistent with the one recently found in Ref.~\cite{Lin2022}, which estimates that the relative volume of the local set to the no-signaling for a fixed number of settings rapidly increases with the number of outcomes. In particular, for two and three measurement settings, they showed that the local set is almost 100\% when $d \ge 6$. 

Let us mention some open problems left from the present work. Clearly, we have only provided numerical evidence of almost certain Bell-violation if we employ all possible Bell inequalities available for the particular Bell scenario relevant to the qutrit, ququart, and ququint case. An obvious question that remains open is whether one can provide an analytic proof for these observations, or a counterexample to the conjecture that one gets almost certain Bell violation in these cases, except for a set of measure zero. Secondly, an observant reader has probably noticed that in many of these visibility distributions presented, one finds a sharp change in the probability amplitude at $v^*=1$. We suspect that this reflects some subtle structure of the set of correlations attainable by maximally entangled states~\cite{Lang_MES}, but the precise nature of this implication on the geometry~\cite{Goh2018,Lin2022} remains to be clarified.

Yet another direction of research would be to better understand the sharp decrease in the probability of violation for $d=7$, even though we once again have access to a complete set of MUBs in this case. A natural guess is that, in this case, considering only the Bell inequalities lifted from $\L_{2;k}$ is no longer sufficient to give a high chance of Bell violation, but it could also be that the correlations themselves are mostly Bell-local. In the case of the former, consideration of Bell inequalities that are naturally suited for maximally entangled states, such as those discussed in~\cite{Salavrakos:PRL:2017,Armin:wp} could be useful. On a closely related note, for the $d=6$, it is also natural to wonder what happens if we make use of a set of approximately MUBs. Our preliminary investigation based on the approximately MUBs of~\cite{Shparlinski:2006wo} and 1500 trials shows that the probability of violation gets boosted to $69.8 \%$  with a 95\% confidence interval of {$[67.41\%, 72.12\%]$}. Since there are also other constructions of approximately MUBs, a systematic investigation is probably best left as some future work. Finally, in analogy to the analysis presented in~\cite{Liang:PRL:2010,FonsecaParisio2015:PRA}, it may also be of interest to understand whether a maximally entangled two-qudit state is, even with the constraint of measuring $d$-dimensional MUBs, the state giving the largest probability of violation. 

\acknowledgements
We would like to thank Cyril Branciard for discussions and two anonymous referees for helpful suggestions. This work is supported by the Ministry of Science and Technology, Taiwan  (Grants No. 107-2112-M-006-005-MY2, 109-2627-M-006-004, 109-2112-M-006-010-MY3), and the National Center for Theoretical Sciences, Taiwan.
 
\bibliographystyle{apsrev4-1}
\bibliography{BIVWithRandomMUBs}

\appendix

\section{Extraction of a two-setting correlation from an $m$-setting correlation}
\label{App_extraction}

Here we illustrate what it means to extract a two-setting correlation from an $m$-setting by a simple example.
Let $\sigma_i \in \{ \sigma_x, \sigma_y, \sigma_z \}$ be the qubit Pauli observables. Consider the three-setting correlation $\vecP$ obtained by Alice measuring the Pauli observables $\sigma_i$  and Bob measuring the rotated Pauli observables $R \sigma_i R^T$, where 
$R = 
\begin{pmatrix}
\cos t & -\sin t \\
\sin t & \cos t 
\end{pmatrix}$.
Written in a tabular form where each $2\times 2$ block corresponds to a fixed pair of $(x,y)$, and with the values of $x$ and $a$ increasing downward while that of $y$ and $b$ increasing rightward:
\begin{align}
   &P(a,b|x,y) = \nonumber \\
&\hspace{0.5em}
\begin{pmatrix}
 \frac{1}{4} \left(c_{2 t}+1\right) & \frac{1}{4} \left(1-c_{2 t}\right) & \frac{1}{4} & \frac{1}{4} & \frac{1}{4} \left(s_{2 t}+1\right) & \frac{1}{4} \left(1-s_{2 t}\right) \\ 
 \frac{1}{4} \left(1-c_{2 t}\right) & \frac{1}{4} \left(c_{2 t}+1\right) & \frac{1}{4} & \frac{1}{4} & \frac{1}{4} \left(1-s_{2 t}\right) & \frac{1}{4} \left(s_{2 t}+1\right) \\
 \frac{1}{4} & \frac{1}{4} & 0 & \frac{1}{2} & \frac{1}{4} & \frac{1}{4} \\
 \frac{1}{4} & \frac{1}{4} & \frac{1}{2} & 0 & \frac{1}{4} & \frac{1}{4} \\
 \frac{1}{4} \left(1-s_{2 t}\right) & \frac{1}{4} \left(s_{2 t}+1\right) & \frac{1}{4} & \frac{1}{4} & \frac{1}{4} \left(c_{2 t}+1\right) & \frac{1}{4} \left(1-c_{2 t}\right) \\
 \frac{1}{4} \left(s_{2 t}+1\right) & \frac{1}{4} \left(1-s_{2 t}\right) & \frac{1}{4} & \frac{1}{4} & \frac{1}{4} \left(1-c_{2 t}\right) & \frac{1}{4} \left(c_{2 t}+1\right) 
\end{pmatrix} \nonumber \\
\end{align}
where $c_{2t} = \cos(2t)$ and $s_{2t} = \sin(2t)$. 
Note that for uniform random $t$, $\vecP$ is a correlation where the qubit MUBs between Alice and Bob are related by the rotation matrix $R$, which is a probability-0 event if they were picking MUBs uniformly at random.
From $\vecP$ we can extract the two-setting correlations $\vecP'_{x_1,x_2;y_1,y_2}$ where $(x_1,x_2), (y_1,y_2)$ are the respective pair of settings for Alice and Bob.
Each $\vecP'$ is just the $4\times 4$ matrix formed by taking the corresponding $2\times 2$ blocks $P(a,b|x_i,y_j), i,j = 1,2$ from $\vecP$. For instance, if we choose $(x_1,x_2)=(1,2)$, then we have
\begin{align}
&P'_{1,2;1,2} = 
\begin{pmatrix}
 \frac{1}{4} \left(c_{2 t}+1\right) & \frac{1}{4} \left(1-c_{2 t}\right) & \frac{1}{4} & \frac{1}{4} \\
 \frac{1}{4} \left(1-c_{2 t}\right) & \frac{1}{4} \left(c_{2 t}+1\right) & \frac{1}{4} & \frac{1}{4} \\
 \frac{1}{4} & \frac{1}{4} & 0 & \frac{1}{2} \\
 \frac{1}{4} & \frac{1}{4} & \frac{1}{2} & 0 \\
\end{pmatrix}, \nonumber \\
&P'_{1,2;1,3} = \nonumber \\
&\hspace{0.5em}
\begin{pmatrix}
 \frac{1}{4} \left(c_{2 t}+1\right) & \frac{1}{4} \left(1-c_{2 t}\right) & \frac{1}{4} \left(s_{2 t}+1\right) & \frac{1}{4} \left(1-s_{2 t}\right) \\
 \frac{1}{4} \left(1-c_{2 t}\right) & \frac{1}{4} \left(c_{2 t}+1\right) & \frac{1}{4} \left(1-s_{2 t}\right) & \frac{1}{4} \left(s_{2 t}+1\right) \\
 \frac{1}{4} & \frac{1}{4} & \frac{1}{4} & \frac{1}{4} \\
 \frac{1}{4} & \frac{1}{4} & \frac{1}{4} & \frac{1}{4} \\
\end{pmatrix}, \nonumber \\
&P'_{1,2;2,3} = 
\begin{pmatrix}
 \frac{1}{4} & \frac{1}{4} & \frac{1}{4} \left(s_{2 t}+1\right) & \frac{1}{4} \left(1-s_{2 t}\right) \\
 \frac{1}{4} & \frac{1}{4} & \frac{1}{4} \left(1-s_{2 t}\right) & \frac{1}{4} \left(s_{2 t}+1\right) \\
 0 & \frac{1}{2} & \frac{1}{4} & \frac{1}{4} \\
 \frac{1}{2} & 0 & \frac{1}{4} & \frac{1}{4} \\
\end{pmatrix}, \nonumber \\
\end{align}
where we have omitted the arguments $(a,b|x,y)$ for brevity.
This procedure generalizes to any correlation $\vecP$ with $m$ settings and $d$ outcomes, where the tabular form of $\vecP$ is an $m \times m$ block matrix with blocks of size $d \times d$, which means the extracted correlations $\vecP'$ will be $2d \times 2d$ submatrices of $\vecP$.

% \section{Visibilities}
% \label{App_visibilities}

% For completeness, we provide here the plots for the white-noise visibility distribution for the case of $d=6$ and $d=7$ (where only up to five MUBs are considered).

\clearpage

\end{document}